\documentstyle[12pt]{article}

\def\be{\begin{equation}}
\def\ee{\end{equation}}
\def\bea{\begin{eqnarray}}
\def\eea{\end{eqnarray}}

\def\p{{\not\! p}}
\def\k{{\not\! k}}
\def\d{\partial}
\def\e{\varepsilon}
\def\g{\gamma}

\def\vec#1{\mbox{\boldmath $#1$}}

\begin{document}

\title{On quantization of massive non-Abelian gauge fields }

\author{Tsuguo MOGAMI
	\thanks{e-mail: mogami@brain.riken.go.jp}\\
	RIKEN, Hirosawa 2-1, Wako-shi,
	Saitama, 351-0198 Japan \\
}

\date{June 30, 2007}

\maketitle

\begin{abstract}
A simpler method of quantization is given for massive gauge theories.  This method gives the same results as those of the conventional massive gauge theory with ghost and Higgs fields under the Higgs mass.  Besides, we point out physical importance of helicity zero states in non-Abelian gauge theories even in massless case.  Furthermore, forms of mass terms that were impossible before, e.g.\ symmetric mass, are possible now.  Applying our method to $SU(2)\times U(1)$ symmetry has no particular difficulty, and gives a variant of the standard model without the Higgs boson.
\end{abstract}

We consider a gauge field theory with the following Lagrangian density.
\be
	{\cal L} = -\frac 1 4 (F_{\mu\nu}^a)^2 - {1\over 2\xi } (\d _\mu A^{a\mu})^2
	+ {m^2\over 2} (A^{a\mu})^2, \label{Lg}
\ee
\[
	F_{\mu \nu } = \d_\mu A_\nu  -  \d_\nu A_\mu  - ig [A_\mu , A_\nu ].
\]
Here $A_\mu^a$ is a $SU(N)$ gauge field, $g$ is a small coupling constant.  We will take the limit of $\xi \rightarrow  \infty $ after renormalization and obtain physical quantities\cite{LY}.  This is different from the unitary gauge, which takes the limit far before renormalization and becomes unrenormalizable.

This theory can be quantized in the ordinary method of canonical quantization.  Four polarization states including a negative norm state will appear.  The negative norm state will have positive energy and large mass.  Let us begin with the canonical variables $\pi ^\mu  = \delta L/\delta 'A_\mu $ and $A_\mu $.  Their commutators are
\be
		[A^a_\mu (\vec x, t), \pi ^{b\nu }(\vec y, t)] = i\delta ^{ab}\delta _\mu ^\nu \delta ^3(\vec x - \vec y),
\ee
and the other commutators are all zero.   After a tedious calculation to obtain eigenstates of the non-interacting part of the Hamiltonian, we have three eigenstates with energy $k_0 = \sqrt {\vec k^2 + m^2}$ and one eigenstate with energy $k'_0 = \sqrt {\vec k^2 + M^2}$ for a 3-momenta $\vec k$, where $M=\sqrt \xi m$.  The commutation relations  for their creation and annihilation operators are
\begin{eqnarray}
	[a_{\sigma}^a (\vec k), {a_{\sigma' }^{b\dagger}(\vec k')}] &=& \delta^{ab} \delta _{\sigma \sigma'}\delta ^3(\vec k - \vec k'), \; (\sigma ,\sigma ' = L,+, -) 	\cr
	[a_{S'}^a(\vec k), a_{S'}^{b\dagger}(\vec k')] &=& -\delta^{ab} \delta ^3(\vec k - \vec k'),
\end{eqnarray}
and zero for the others.  The definition of the vacuum is $a_{\sigma}(\vec k)|0\rangle= 0 \; (\sigma = S', L, +, -)$.  The state created by $a_{S'}^\dagger (\vec k)$ is a negative norm state and will be called ``scalar polarization state" or ``$S'$ state".  Please note that the energy of this state $\sqrt {\vec k^2 + M^2}$ is positive, and this state have a polarization vector $k'_\mu /M = (k'_0, \vec k)/M$.  The other polarization states have polarization vectors that are orthogonal to $k_\mu = (k_0, \vec k)$.  There is no reason to consider those three states to be unphysical since they have positive norm and a positive energy.

The field $A_\mu $ may be expressed in terms of these creation and annihilation operators as
\be
	A^a_\mu (x) = \int {d^3k \over (2\pi)^{3/2}}
	\left\{\sum_{\sigma =L, +, -} a^a_{\sigma }(\vec k)
	{e^{-ikx} \over \sqrt {2k_0}} \e ^\sigma _\mu (k) 
	+ a^a_{S'}(\vec k)  {e^{-ik'x} \over\sqrt {2k'_0}} \sqrt \xi 
	\; \e ^{S'}_\mu(k')  + {\rm c.c.} \right\}. \label{Aexp}
\ee
Here, $\e ^\sigma $'s are polarization vectors and
\be
\begin{array}{rl}
	\e ^\sigma(k) \cdot \e ^{\sigma '}(k) & = -\delta _{\sigma \sigma'},\\
	\e ^\sigma(k)  \cdot \e ^S(k) & = 0, \; (\sigma ,\sigma ' = L, +, -)  \\
	\e ^S(k) \cdot \e ^S(k) & = 1.
\end{array}
\ee
We chose the spatial three vector of $\e ^L_\mu $ to be proportional to $\vec k$.  If we write down their explicit form when $\vec k$ is in $1$ direction,
\be
\begin{array}{c}
	\e ^S_\mu  = (k_0, k, 0, 0)/m,  \\
	\e ^L_\mu  = (k, k_0, 0, 0)/m,  \\
	\e ^+_\mu  = (0, 0, 1,+i)/\sqrt 2,  \\
	\e ^-_\mu  = (0, 0, 1, -i)/\sqrt 2.
\end{array}
\ee
The polarization represented by $\e ^+_\mu $ will be called ``longitudinal polarization" or ``$L$ polarization" here\footnote{Please be careful that some of the literatures use ``longitudinal" to mean scalar polarization in this letter.}.  $\e ^+$ and $\e ^-$  are said to have ``transverse" polarization.  Here we defined ``$S$ state" to be a state having mass $m$ and polarization vector $k_\mu /m$ for the convenience of later calculation, even though canonical quantization of our theory (\ref{Lg}) gives $S'$ state and not $S'$ state.  Please note that we have $S'$ state and not $S$ state in the expansion of $A_\mu $ (\ref{Aexp}) too.

The propagator of our gauge field is
\be
	D_{\mu \nu }(k) = {-i \over k^2-m^2}   (g_{\mu \nu }-{k_\mu k_\nu \over m^2})
	+ {k_\mu k_\nu \over m^2} {-i \over k^2-M^2}.\label{prop}
\ee
This propagator propagates $S'$ state with mass $M$, but not $S$ state with mass $m$, following the canonical quantization above.

This theory satisfies the following basic physical requirements.  First, this theory is simply renormalizable
since it has local operators of dimension less than or equal to four\cite{BPHZ}.  Since its S-matrix is matrix elements of $e^{-iHT}$, its unitarity is obvious.  Because the particle with negative norm ($S'$ state) has large mass  $M \equiv \sqrt\xi m$, negative norm state will not appear below this energy level.  This theory will contain only positive norm particles after taking $\xi \rightarrow  \infty $ limit.


On the other hand, the conventional massive gauge theory, which should be compared to our theory, is a gauge theory that acquires mass with Higgs mechanism.  Let us consider Higgs-Kibble model for instance.  Its Lagrangian density is
\be
	{\cal L}_{\rm HK} = -\frac 1 4 (F^a_{\mu \nu })^2
		+ |D_\mu \Phi|^2 
		- \frac \lambda 2 (\Phi^\dagger \Phi - |v|^2)^2,
		\label{HK}
\ee
where its gauge group is $SU(2)$, and $D_\mu $ is covariant derivative.  The field $\Phi$ is a two-dimensional complex field belonging to $SU(2)$ doublet, and has vacuum expectation value $v$.  It is convenient to parameterize $\Phi$ with real scalar fields:
\be
	\Phi(x) = {1\over \sqrt 2}(\chi ^2(x) + i\chi ^1(x), 
		v + \psi (x) - i\chi ^3(x))^T .   \label{Hcmp}
\ee
Here we add gauge-fixing terms:
\bea
	{\cal L}_{\rm gf} &=& - (1/2\xi ) (\d ^\mu A_\mu ^a 
	+ \xi m\chi ^a) ^2 \cr
	&&+ i\ \bar c^a [\d ^\mu D_\mu^{ab} +\xi m^2 \delta^{ab} 
	+ (g/2)\xi m(\psi \delta^{ab}+ f_{acb}\chi^c )] c^b,
\eea
where the numbers $f_{abc}$ are structure constants.  This way of gauge-fixing is called ``$R_\xi$  gauge"\cite{Rxi}.  The fields $\bar c$ and $c$ are Grassmann fields, which are called Faddeev-Popov\cite{FP} ghosts.  Using (\ref{Hcmp}), we may write the free part of the Lagrangian as
\bea
	{\cal L}_{\rm HK}+L_{\rm gf} &=& - \frac 1 4 (F_{\mu \nu }^a)^2 - {1\over 2\xi } (\d _\mu  A^{a\mu })^2
	+ {m^2\over 2} (A^a_\mu )^2 \nonumber\\
	&&+ \frac 1 2 (\d ^\mu \chi )(D_\mu \chi ) 
	- \frac 1 2 M^2 \chi ^2 \cr
	&&+ \frac 1 2 (\d _\mu \psi )^2 
	- \frac 1 2 \mu_H^2 \psi ^2 
	+ \frac g 2 A_\mu ^a (\chi ^a\d ^\mu \psi  
	- \psi \d ^\mu \chi ^a)    \nonumber\\
	&&+ i\ \bar c \ \d _\mu (D_\mu  c) + M^2 \bar c c,\label{LHK}
\eea
where the interaction terms that is linear in $A_\mu ^a$ are kept for the purpose of later 1-loop calculation.  We see that the gauge fields, Higgs $\chi $, Higgs $\psi $ and the ghost fields acquire mass $m (= gv/2), M (= \sqrt \xi m),  \mu _H (= \sqrt \lambda v)$ and $M$ respectively.

Our theory gives the same physical results as that of the conventional Higgsed gauge theory  when the energy is lower than the Higgs masses ($E \ll  \mu_H, E \ll M$).  That is because the difference between the two theories comes from Higgs and ghost fields and the effects of these heavy fields are small for lower energy phenomena.

Here we introduce an easier method of calculating amplitudes, which gives the same result as using propagator (\ref{prop}) and ordinary Feynman rules.  Repeating small gauge transformation $A_\mu (x) \rightarrow  A_\mu (x) + N^{-1} D_\mu \phi (x)/m$ for $N$ times and taking $N\rightarrow  \infty $ limit, $A_\mu = T^a A^a_\mu $ is transformed into
\begin{eqnarray}
	A_\mu  &\rightarrow & e^{ig\phi /m} ( i\d _\mu /g + 	A_\mu ) e^{- ig\phi /m} \nonumber\\
	&=& A_\mu  + \d _\mu \phi /m + ig [\phi /m, A_\mu ] + \cdots \equiv  A'_\mu ,\label{fnG}
\end{eqnarray}
where $T^a$'s are generators of the symmetry group.  The original Lagrangian (\ref{Lg}) may be rewritten as
\[
	{\cal L} = - \frac 1 4 (F_{\mu \nu }^a)^2  
		+ {m^2 \over 2} (A^a_\mu )^2
		- m A^a_\mu \d ^\mu \phi^a + \frac 1 2 \xi m^2 \phi ^2
\]
by introducing a scalar auxiliary field $\phi^a $.  Applying transformation (\ref{fnG}), our Lagrangian is transformed into\footnote{ The higher order interaction terms of $\phi $ in $L_\phi $ are unrenormalizable.  The unrenormalizable contributions to the Green functions from these terms will precisely be canceled with the unrenormalizable contributions from the Proca field, because we know that the equivalent original theory is renormalizable from the beginning.}
\bea
	{\cal L} &\rightarrow &  - {1\over 4} (F_{\mu \nu }^a)^2 + {m^2\over 2} (A^a_\mu )^2 \nonumber\\
	&& - {1\over 2} (\d ^\mu \phi ) (D_\mu \phi ) + {1\over 2} M^2 \phi ^2
	+ (\hbox{higher order terms in } \phi ) \equiv  {\cal L}_\phi .
\eea
To calculate n-point function
\be
	\int {\cal D}A\  e^{i\int d^4 x {\cal L}(x)}  A_\mu (x_1) A_\nu (x_2) \cdots 
	\equiv \langle A_\mu (x_1) A_\nu (x_2) \cdots\rangle_{A, \xi }, \label{e1}
\ee
we use the fact that it is equivalent to
\be
	= \int {\cal D}\phi {\cal D}A\  e^{ i\int d^4 x {\cal L}_\phi (x) }  A'_\mu (x_1) A'_\nu (x_2) \cdots 
	\equiv \langle A'_\mu (x_1) A'_\nu (x_2) \cdots\rangle_{\phi , A, \xi } \label{e2}
\ee
using transformation (\ref{fnG}).  In the latter form, $A_\mu $ is now a Proca field and its propagator is
\be
	D'_{\mu \nu }(k)   = \frac{-i}{k^2-m^2} (g_{\mu \nu }-{k_\mu k_\nu \over m^2} ),	\label{prop2}
\ee
which does not propagate $S'$ state.  The effects of $S'$ states are replaced by those of the scalar field $\phi $, since the two formula (\ref{e1}) and (\ref{e2}) should be equivalent.  This equivalence can also be proven by repeatedly applying identity (\ref{WT}) to the Feynman diagrams of the original theory (\ref{e1}).

Now, let us check that $S'$ states do not give any contribution to our theory in the limit $\xi \rightarrow  \infty $ at 1-loop level.  As an example, two point function
\be
	\langle A_\mu (x_1) A_\nu (x_2) \rangle_{A, \xi }
	=\langle A'_\mu (x_1) A'_\nu (x_2)\rangle_{\phi , A, \xi }	\label{2pt}
\ee
will be considered.  We need, however, only the physical polarizations, and scalar polarized part $\langle \d _\mu \phi (x) A'_\nu (0) \rangle /m$ will be discarded later even if we calculate it.  In general, because any physical quantity such as S-matrix is computed from the gauge invariant combinations of $A_\mu $, the difference between $A'_\mu $ and $A_\mu $, i.e.\ $\d _\mu \phi /m + i [\phi /m, A_\mu ]  +\cdots$, does not give any physical effect.  Then, we only have to consider
\be
	\langle A_\mu (x_1) A_\nu (x_2)\rangle_{\phi , A, \xi }.
\ee
The lowest nontrivial order contributions to this two-point function are the graphs shown in fig.1.  (One can directly check that the scalar loop in fig.1 is equivalent to the looping of a $S'$ state in (\ref{e1}) by applying identity (\ref{WT}).)  This scalar loop gives
\be
	{ig^2 \over 2 } f_{abc} f_{a'bc}\int {d^4k\over(2\pi )^4}
	\frac 1 4 (q-2k)_\mu (q-2k)_\nu  
	{i\over k^2-M^2} {i\over  (q-k)^2-M^2},\label{1Lp}
\ee	
and the momentum integration gives
\be
	= ig^2 f_{abc} f_{a'bc} (q^2g_{\mu \nu }- q_\mu q_\nu ) 
	\left\{- \log \Lambda 
	+ (1-{4 M^2\over q^2})^{3/2} {\rm arccot}\sqrt {1-{4 M^2\over q^2}}\right\}
\ee
abbreviating the constant factors, and $\Lambda$ is the cutoff.  Only a small contribution of $q^2 \, O(q^2/M^2)$ is left after renormalization for $ q^2\ll  M^2$, and it goes to zero in the limit of $\xi \rightarrow \infty $.

Next, Let us check that our theory gives the same result as that of the conventional Higgsed theory.  Our theory removes $S'$ states by making its mass $M$ very large.  On the other hand, the conventional theory removes $S'$ states by using ghost and Higgs particles.  The action of ghost fields takes the form of
\[
	+ i \bar c [\d ^\mu D_\mu + M^2] c
\]
as is shown in (\ref{LHK}).  This is apparently same as the action of $\phi $ in ${\cal L}_\phi $
\[
	\frac 1 2 \phi [\d ^\mu D_\mu + M^2]\phi ,
\]
and the ghosts give just $-2$ times of the scalar loop (\ref{1Lp}) to the two point function (\ref{2pt}), when the unphysical $q_\mu$  proportional part is ignored.  Where, the negative sign comes from anticommutation of the Grassmann field.  Finally, the action for Higgs field is
\[
	- \frac 1 2 \chi [\d ^\mu D_\mu + M^2]\chi ,
\]
and gives the same\footnote{
The loop of the negative norm states has the same sign as that of positive norm scalar loop.  Please do not confuse having negative norm with having negative sign for a loop.} loop integration as (\ref{1Lp}).
These contributions add up to completely cancel contributions from $S'$ states.  This cancellation is said to be necessary to keep unitarity of S-matrix.

This fact tells us the reason why it was believed that Higgs mechanism was indispensable for massive gauge bosons.  A loop of the $S'$ state has a symmetry factor $1/2$, because it is effectively a real scalar field $\phi $.   On the other hand, the ghost loop has factor $-1$, which is twice as much as needed.  This excessive correction is inevitable because anti-commuting field cannot have kinetic term of the form $(\d _\mu c)(\d _\mu  c)$, but only $(\d _\mu \bar c)(\d _\mu  c)$ is possible.  To eliminate this overcorrection, another unphysical field of mass $M$ is required.  That is the Higgs field.

Even in the limit of $m \rightarrow 0$, the ghosts give twice excessive correction over the effect from negative norm states (i.e.\ $S'$ states).  What does it eliminate?  Originally, ghost fields are introduced as a trick to make only picking transverse polarizations up consistent\cite{FyG}.
Then the ghosts should be canceling not only $S'$ states but also $L$ states.

Actually, the amplitude of scattering of particles giving rise to final state that include $L$ state does not vanish even in $m \rightarrow  0$ limit.  Let us directly demonstrate it with the example of fermion-anitifermion scattering process, which is shown in fig.2.  The amplitude is
\bea
	i M^{\mu \nu }&=&(ig)^2 \bar v(p_+) \{\g ^\mu T^a  {1\over \p-\k_2-m_{\rm f}} \g ^\nu  T^b 
	+\g ^\nu  T^b {1\over \k_2-\p_+ -m_{\rm f}}\g ^\mu  T^a \} u(p) \cr
	&&+ ig^2 \bar v(p_+) \g _{\rho'} T^c u(p) \times D_{\rho '\rho }(k_3) f_{abc} V^{\mu \nu \rho },
\eea
where $m_{\rm f}$ is fermion mass, the antifermion has momentum $p_+$ and spinor $\bar v(p_+)$, the other fermion has $p$ and $u(p)$, momentum conservation requires $p_+ + p = - k_3, k_1+k_2+k_3 = 0$, and
\be
	V_{\mu \nu \rho } = g_{\mu \nu }(k_2-k_1)_\rho + g_{\nu \rho }(k_3-k_2)_\mu + g_{\rho \mu }(k_1-k_3)_\nu .
\ee
Let us consider the case where the polarization of the second boson in the final state is $\e ^*_\nu (k_2) = k_{2\nu }/m$.  The amplitude gets
\bea
	 i M^{\mu \nu } \e ^*_\mu (k_1) \e ^*_\nu (k_2) 
	&=& - ig^2 \e ^*_\mu (k_1) m^{-1} f_{abc}
	(k_1^2 g^{\mu \rho }- k_1^\mu k_1^\rho - m^2 g^{\mu \rho })\cr
	&&\times  D_{\rho '\rho }(k_3) \bar v(p_+)  T_c \g _\rho u(p), 
\eea
where we have used
\be
	V_{\mu \nu \rho } k_2^\nu 
	= - (k_1^2 g_{\rho \mu } -k_{1\rho }k_{1\mu })
	+ (k_3^2 g_{\rho \mu }-k_{3\rho }k_{3\mu }),\label{WT}
\ee
and from Dirac equations for final state spinors $u, \bar v$,
\be
	k_{3\rho }\bar v  T^c \g ^\rho u = - \bar v(p_+)  T^c [(\p_+ +m_{\rm f}) + (\p - m_{\rm f})] u(p) = 0.
\ee
When the other boson has $L$ or transverse polarization\footnote{ This vanishing occurs even in higher order in perturbation because this fact follows conservation of the current.  And, this holds for every mass for $k_2$ because neither $k_2^2 = m^2$ nor $k_2^2 = M^2$ was used.}
 ($\e _1\cdot k_1 =0,\  k_1^2 =m^2$),
\be
	i M^{\mu \nu } \e ^*_{\mu }(k_1) k_{2\nu } = 0.
\ee
When the other boson has $\e ^*_\mu (k_1) = k_{1\mu }/m$ polarization we have,
\be
	i M^{\mu \nu }  k_{1\mu } k_{2\nu }/m^2 = 1,\label{SS}
\ee
and finite in the $m \rightarrow  0$ limit.  Still, $S$ states will not appear in our theory, since our propagator (\ref{prop}) does not include $S$ polarization of mass $m$, and $S'$ states are too heavy to appear.  Now let us calculate the amplitude for two longitudinal final bosons.  For calculation, we define
\[
	\e ^B_\mu (k) \equiv  \e ^L_\mu (k) - {k_\mu \over m}
	= {m \over |\vec k|+\sqrt {\vec k^2+m^2} } (1, - {\vec k \over |\vec k|}).
\]
With this, we have
\be
	\e ^L_\mu (k_1) \e ^L_\nu (k_2) = -k_{1\mu } k_{2\nu }/m^2
	+ \e ^L_\mu(k_1) k_{2\nu }/m +k_{1\mu } \e^L _\nu (k_2)/m + \e ^B_\mu (k_1) \e ^B_\nu (k_2).
\ee
We already calculated that the second and the third term is zero, and the fourth term is negligible in $m \rightarrow  0$ limit, since $\e ^B$ is proportional to m.  Then we get
\be
	M^{\mu \nu } \e ^{L*}_\mu(k_1) \e ^{L*}_\nu(k_2)  
	=  - M^{\mu \nu } k_{1\mu }k_{2\nu }/m^2,
\ee
and the amplitude of the final state being two $L$-polarized bosons is finite and equal to (\ref{SS}) in the $m \rightarrow  0$ limit.  (Note that the probability of having $L$ state particles is finite as $m \rightarrow  0$, since $(\e ^L)^2$ keeps constant even though elements of $\e ^L_\mu (k)$ get infinitely large in this limit.)  Now we see that the ghosts are eliminating this appearance of $L$ states too by that excessive factor $2$.

Only when $m = 0$, it is possible to remove not only $S'$ states but also $L$ states from the theory.  This is the reason why it was possible only for massless cases to quantize gauge theories using ghosts.
Let us see it in an actual calculation.  For the states with momentum $(\sqrt {m^2+k_3^2}, 0, 0, k_3)$, the transverse polarization vectors are $(0, 1, 0, 0)$ and $(0, 0, 1, 0)$.  If we boost these state by $k_1/\sqrt {m^2+k_3^2}$ into $1$ direction, their momentum gets $(\sqrt {m^2+k_1^2+k_3^2}, k_1, 0, k_3)$.  Then the polarization vector $(0, 1, 0, 0)$ is boosted into
\[
	\e = ({k_1\over \sqrt {m^2+k_3^2}}, 
	{\sqrt {m^2+\vec k^2} \over \sqrt {m^2+k_3^2}},
	 0, 0),
\]
where $\vec k = (k_1, 0, k_3)$.  This is not transversely polarized in this Lorenz frame since its spatial component is not orthogonal to $\vec k$.  Using the longitudinal and transverse polarization vectors of this frame
\bea
	\e^L(k) &=& \frac 1 m (|\vec k|, \sqrt {m^2+\vec k^2} {\vec k \over |\vec k|}), \cr
	\e^T(k) &=& {1\over |\vec k|} (0, k_3, 0, -k_1), 
	\nonumber 
\eea
vector $\e$ is decomposed as
\be
	\e = {m k_1 \over |\vec k| \sqrt {m^2+k_3^2}} \e^L 
	+ {\sqrt{m^2+\vec k^2} \over \sqrt{m^2+k_3^2}} \e^T. 
	\label{mix}
\ee
Thus, when the theory is massive, transverse polarization will be mixed with $L$ polarization by boosting, and then we cannot make covariant theory without $L$ states.  Only when the theory is massless, it is possible to remove $L$ states by hand, because the coefficient of $\e^L$ in (\ref{mix}) goes to zero.   Then the conventional gauge theory with ghosts is consistent despite its excessive correction that eliminates $L$ states.

Now, we see that two different kinds of gauge theories are possible when $m = 0$.  Both of the theories satisfy unitarity and have positive norms and positivity of energy.  One kind is the conventional massless theories with ghosts only.  The other is $m \rightarrow  0$ limit of our theory, which includes $L$ states with finite probability.  Taking $m \rightarrow  0$ limit in the conventional Higgsed gauge theories with fixed Higgs mass gives the same as the latter.  The former are isolated theories, i.e.\ they are possible only when $m = 0$.  The latter are possible for general $m$ including $m \rightarrow  0$ limit.  Since both satisfy physical requirements, determining which is correct needs experimentation.

It has long been thought that proving quark confinement in QCD is very difficult.  It is possible that this difficulty comes from choosing wrong one of the two theories.

We think that lattice QCD agrees with our theory, because the continuum limit in the lattice theories is essentially taking $m \rightarrow 0$ limit for massive theories on the lattice.

$L$ polarization states may be restated as helicity$=0$ states.   We now know that these states have finite probability.  Therefore, even in such theories as gravity and supersymmetry, we should consider adopting other helicity states than the highest and lowest helicity states.

Furthermore, the forms of mass terms that were impossible before are possible in our theory.  For example, all the gauge bosons in a representation of $SU(N)$ could not have the same mass except for $N=2$ in the conventional theories.
To restore the physical freedoms which were removed by the ghost fields, we need $N^2-1$ Higgs bosons.  Since Higgs field in the fundamental representation has only $2N$ real freedoms, it is impossible for all the gauge bosons to have mass except for $N=2$ (Higgs-Kibble model).  If the Higgs field is in adjoint representation, the gauge boson related to the unbroken generator will remain massless.  Our theory, however, does not have such limitation related to Higgs mechanism.

If we apply our method to the electroweak theory\cite{GWS},
we obtain a renormalizable theory, which gives the same physical results as the conventional Higgsed theory under the Higgs mass.  Our Lagrangian is
\bea
	{\cal L}_{\rm std} &=& - \frac 1 4 (\d _\mu A_\nu  - \d _\nu A_\mu  - ig [A_\mu , A_\nu ] )^2
	-\frac 1 4 (\d _\mu B_\nu  - \d _\nu B_\mu )^2\label{Lstd}\\
	&&- \frac 1 {2\xi } (\d _\mu  A^{a\mu })^2 
	- \frac 1 {2\xi } (\d ^\mu  B_\mu )^2 
	+ \frac 1 2 M_W^2 ((A^1_\mu )^2 + (A^2_\mu )^2 
	+ (A^3_\mu - \tan \theta _W B_\mu )^2), \nonumber
\eea
where $A^a_\mu $ and $B_\mu $ are $SU(2)$ and $U(1)_Y$ gauge fields  respectively, $M_W$ is weak boson mass, and $\theta _W$ is Weinberg angle.  Unitarity and renormalizability are obvious in this theory.

We don't have negative norm states in this theory (\ref{Lstd}).
Let us call the unbroken $U(1)$ gauge symmetry $U(1)_\g $ and the corresponding gauge boson ``photon".  The $S'$ polarization states of other gauge bosons will not have any effect, because they are simply heavy as is the case of Higgs-Kibble model.
Finally, the photon $S'$ states and $L$ states will be neither emitted nor absorbed as they didn't in QED, because the breaking of $U(1)_\g $ is proportional to $1/\xi $, and negligible when $\xi $ is very large.

This theory (\ref{Lstd}) agrees with the conventional Higgsed electroweak theory when the energy level is much lower than the Higgs mass $\mu _H$ and $\xi $ is very large.
The difference between these two theories in the $R_\xi $ gauge is the ghost and Higgs fields.  The Higgs fields and heavier ghosts will not appear because of their heavy mass $\sqrt \xi M_Z, \sqrt \xi M_W$ and $\mu _H$.
The massless ghost corresponding to photon will be cancelled with the $S'$ states and $L$ states owing to BRS symmetry\cite{BRS} related to $U(1)_\g $.  Therefore, the two theories agree in that $S'$ and $L$ states will have no physical effect.

Finally, let us consider the letponic part of the standard model.
The trouble is that the fermion mass terms explicitly break $SU(2)\times U(1)$ symmetry.  In the conventional understanding, BRS symmetry should be conserved to prevent appearance of negative norm states.  Then, consistent theories are obtained\cite{tV} only when we introduced mass terms via the vacuum expectation value of the Higgs field.  On the other hand, our theory don't need such mechanism and may simply introduce fermion mass terms, because the negative norm states are ruled out by simply giving unphysically large mass.

\section*{Figure Legends}

\noindent {\bf Figure1.} The 1-loop graphs contributing to the two-point function.  The wavy lines of the gauge bosons represent the propagator in (\ref{prop2}) here.

\noindent {\bf Figure2.}  The tree-level diagrams contributing to the fermion-antifermion scattering into two gauge bosons.

\end{document}